\documentclass[journal=jacsat,manuscript=article]{achemso}

\usepackage{chemformula} 
\usepackage[T1]{fontenc} 
\usepackage[abbreviations=false,%
			detect-all,%
			list-units=single,%
			per-mode=reciprocal,%
			exponent-product = \cdot]{siunitx}
\usepackage{anyfontsize}
\usepackage{physics, mathtools}
\usepackage{hyperref}

\author{Dmitry Nuzhdin}
\affiliation{European Laboratory for Non-linear Spectroscopy (LENS), University of Florence, 50019 Sesto Fiorentino, Italy}
\alsoaffiliation{Institute of Applied Physics, Karlsruhe Institute of Technology (KIT), 76128 Karlsruhe, Germany}
\altaffiliation{Current address: Aix Marseille Univ, CNRS, Centrale Marseille, Institut Fresnel, F-13013 Marseille, France}
\email{nuzhdin@fresnel.fr}
\phone{+33 (0)4 91 28 83 28}
\fax{+33 (0)4 91 28 80 67 }
\author{Lorenzo Pattelli}
\affiliation{European Laboratory for Non-linear Spectroscopy (LENS), University of Florence, 50019 Sesto Fiorentino, Italy}
\alsoaffiliation{Istituto Nazionale di Ricerca Metrologica (INRiM), 10135 Torino, Italy}
\author{Sara Nocentini}
\affiliation{European Laboratory for Non-linear Spectroscopy (LENS), University of Florence, 50019 Sesto Fiorentino, Italy}
\alsoaffiliation{Istituto Nazionale di Ricerca Metrologica (INRiM), 10135 Torino, Italy}
\author{Diederik Wiersma}
\affiliation{European Laboratory for Non-linear Spectroscopy (LENS), University of Florence, 50019 Sesto Fiorentino, Italy}
\alsoaffiliation{Consiglio Nazionale delle Ricerche - Istituto Nazionale di Ottica, 50019 Sesto Fiorentino, Italy}
\alsoaffiliation{Istituto Nazionale di Ricerca Metrologica (INRiM), 10135 Torino, Italy}
\email{wiersma@lens.unifi.it}
\phone{+39 (0) 05 54 57 25 17}
\fax{+39 (0) 05 54 57 49 32}

\title[An \textsf{achemso} demo]
  {Diagnostics and characterization of photonic circuits by wide-field spatio-temporal imaging}


\begin{document}

\begin{abstract}

The growing diffusion of integrated photonic technologies requires fast and non-invasive quality control techniques for mass-production. We present a general diagnostic technique for sub-ps imaging of photonic circuits combining wide-field optical microscopy and optical gating. The simultaneous access to multiple parameters of a photonic structure enables an unprecedented characterization of its functional design as opposed to typical single-domain techniques such as frequency or time domain reflectometry and near-field microscopy. The non-contact and non-perturbative nature of the technique makes it relevant for both planar and three-dimensional circuits, as well as for silicon, polymeric or hybrid platforms. We apply our technique to different photonic chip components fabricated by Direct Laser Writing, revealing the spatial and temporal hallmarks of fabrication imperfections causing losses or deviations from the intended device behavior. At the same time, the technique allows in situ probing of key properties of photonic devices as the local propagation constants of guided modes or the quality factor of resonant elements. Our method is relevant for both the scientific and the industrial communities as it lends itself to be scaled up to in-line quality control thanks to its non-scanning nature.\\

\textit{This document is the unedited Author\textquotesingle s version of a Submitted Work that was subsequently accepted for publication in ACS Photonics, copyright \textcopyright American Chemical Society after peer review. The final edited and published work is available at  \href{https://dx.doi.org/10.1021/acsphotonics.0c00271}{ DOI: 10.1021/acsphotonics.0c00271}.}
\end{abstract}

\section{Introduction}
Rapid progress in lithographic techniques and hybrid material integration is leading to complex on-a-chip platforms with thousands of elements miniaturized into integrated photonic circuits. Nowadays, photonic circuits are widening their applicability thanks to their competitive performances, replacing conventional microelectronics in specific fields ranging from signal processing \cite{Marin-Palomo2017,Willner2014,Tanabe2005,Koos2009} to machine learning \cite{Shen2017, Harris2017, Tait2017} as they represent the convenient platform both at the microscale and towards a world-wide all-optical communication system \cite{moughames2019dimensional}. However, the field still suffers from the lack of adequate diagnostics tools for the functional testing of fabricated devices. Despite the high resolution of well-established lithographic techniques enabling high-fidelity fabrication of designed photonic structures, minor deviations or fabrication defects are inevitable and can strongly affect light propagation even with no apparent deviation from the intended design. At the same time, due to their small scale and irregular geometry, these deviations are both difficult to model and to inspect \cite{Ferrari2009, Melati2014, Faggiani2016}.

In traditional fiber optics technology, high-precision quality control over long propagation distances is obtained through several well-established techniques, which include optical frequency domain reflectometry (OFDR) \cite{Soller2005}, optical time domain reflectometry (OTDR) \cite{Barnoski1977} and their variants \cite{Rogers1981, Ahn2006, VonDerWeid1997, Golubovic1997, Husdi2004}. These approaches allow to investigate the quality of single mode fibers with defect positioning precision of tens of \si{\micro\meter} over tens of meters of total fiber length \cite{Soller2005}. A coherent frequency domain reflectometry was also implemented to study integrated waveguides \cite{Glombitza1993, Yuksel2009, Bru:18} setting the current state-of-the-art in defect positioning resolution slightly above \SI{5}{\micro\meter}, which however remains insufficient to probe photonic structures characterized by sub-wavelength features. Another drawback of reflectometry techniques is the ambiguity in location of defect's position in case of parallel or branching light pathways as in beam splitters or Mach-Zehnder interferometers \cite{Barnoski1977}, as well as the impossibility to distinguish multiple time-coincident reflection peaks. Similarly, more complex systems including, e.g., coupled resonator waveguides \cite{Ferrari2009, Melloni2003}, add-drop filters \cite{Schwelb2004}, complex waveguide networks \cite{Shen2017} or multimode interference devices \cite{Besse1994, Soldano1995}, require more sophisticated investigation techniques to uniquely attribute reflection peaks to a single position in a multi-branched photonic circuit configuration. Moreover, if more complex integrated system are taken in analysis, the reflection at the different ports or in presence of defects can create intermixing terms that introduce an ambiguity in the defect mapping  \cite{Yuksel2009, Munoz2017}.  

Among possible approaches to monitor on-chip photonic platforms, far-field and near-field techniques offer a diverse insight into the sample. Far-field methods typically consist of simple widefield microscopy of the investigated sample exploiting for instance, far-field scattering microscopy (FScM) applied to grated waveguides \cite{Hopman2007} or to photonic crystal structures \cite{Loncar2002}. Herein, the main limitation is provided by the fact that a static characterization cannot reveal the occurrence of multiple signals or revivals repeating at different times in the same location. Additionally, a crucial aspect is the out-coupled light intensity that must be high enough to be detected over the background signal. On the other hand, near-field methods allow sub-wavelength imaging of a photonic device by collecting the evanescent field through a scanning tip, but the perturbation introduced by the probe tip and the long scanning times \cite{Burresi2011} hinder a widespread adoption of near-field techniques for circuit characterization in mass production processes \cite{Balistreri1999, Balistreri2001, Gersen2004}. Additionally, scanning near-field optical microscopy (SNOM) techniques are limited to the study of planar or quasi-2D structures due to their near-field detection, while 3D photonic circuits with stacked components are increasingly gaining momentum \cite{Nesic2019, Nocentini2018}. As an alternative to near-field microscopy, ultrafast photomodulation spectroscopy has also been recently proposed \cite{Bruck2014}. In this case, the spatial mapping of the mode distribution of the device under test is achieved by a point-by-point optical pumping that locally perturbs the refractive index of silicon structures. By monitoring the transmission through the device, a space and time dependent photomodulation map in silicon photonic circuits can be retrieved. However, this method is material-specific and it still requires to scan a pump beam over the sample area.

In this paper, we present a multi-domain approach for the diagnostic of integrated photonic circuits. By combining principles of time-resolved optical gating and wide-field microscopy, we demonstrate far-field temporally and spatially resolved signal tracing with sub-picosecond temporal resolution and $\sim$\si{\micro\meter} spatial resolution. The high temporal resolution allows to detect not only geometrical defects but also functional deviations with respect to the desired operation of photonic structures (e.g., due to density inhomogeneities). It is interesting to comment in more detail on the  similarities and differences between the proposed method and other established reflectometry techniques such as OFDR. In an OFDR measurement, even in the simplest single arm geometry, the presence of multiple reflections can often lead to additional intermixing terms that do not correspond to real reflections, introducing an ambiguity in the interpretation of the resulting spectrum \cite{Yuksel2009, Munoz2017}. In contrast, ultra-fast imaging of PICs allows to decouple in space even two or more events occurring at the same delay in the time domain. Additionally, examining a multiport device with OFDR, one must address individually each combination of output and input ports, which are instead interrogated in parallel in our multi-domain approach.  Finally, on a practical level, any broadband OFDR approach is inherently limited by the need of robust group velocity dispersion compensation \cite{Zhao2017, Melati2016} or by the need of additional arms in the device under test as in the case of sweep-wavelength homodyne interferometric detection \cite{Bru:18}. Arguably, these factors are the reason why the highest defect positioning resolution reported so far is limited to \(5.3 \pm 1.7\)
\si{\micro\meter} \cite{Zhao2017} and more typically in the 10 $\sim$\si{\micro\meter} range \cite{Melati2016, Soller2005, Bru:18}. This value is still far from the diffraction limited resolution for the given set of optics and wavelength for widefield imaging, which is used in the proposed approach. When complemented with subpicosecond temporal resolution of the system, this allows to identify always and unambiguously the origin of each peak, even temporally coincident ones, free of any spurious intermixing terms. On the other hand, when it comes to simpler circuits, OFDR has the advantage of providing both amplitude and phase information, as well as a simpler setup with no moving parts.

Nonetheless, our proposed approach does not rely on specific material characteristics which makes it equally suitable to study dielectric, semiconductor or polymer devices. The spatio-temporal imaging enables also a quantitative characterization of the optical properties of different photonic devices, as we illustratively show for single-mode polymeric waveguides assisted by Bragg-couplers, whispering gallery-mode (WGM) ring resonators vertically coupled to a bus waveguide and complex inter-coupled waveguide-based networks.

\section{Results and discussion}
\subsection{Principles of ultra-fast spatio-temporal imaging}
\begin{figure}[ht]
\centering\includegraphics[width=\textwidth]{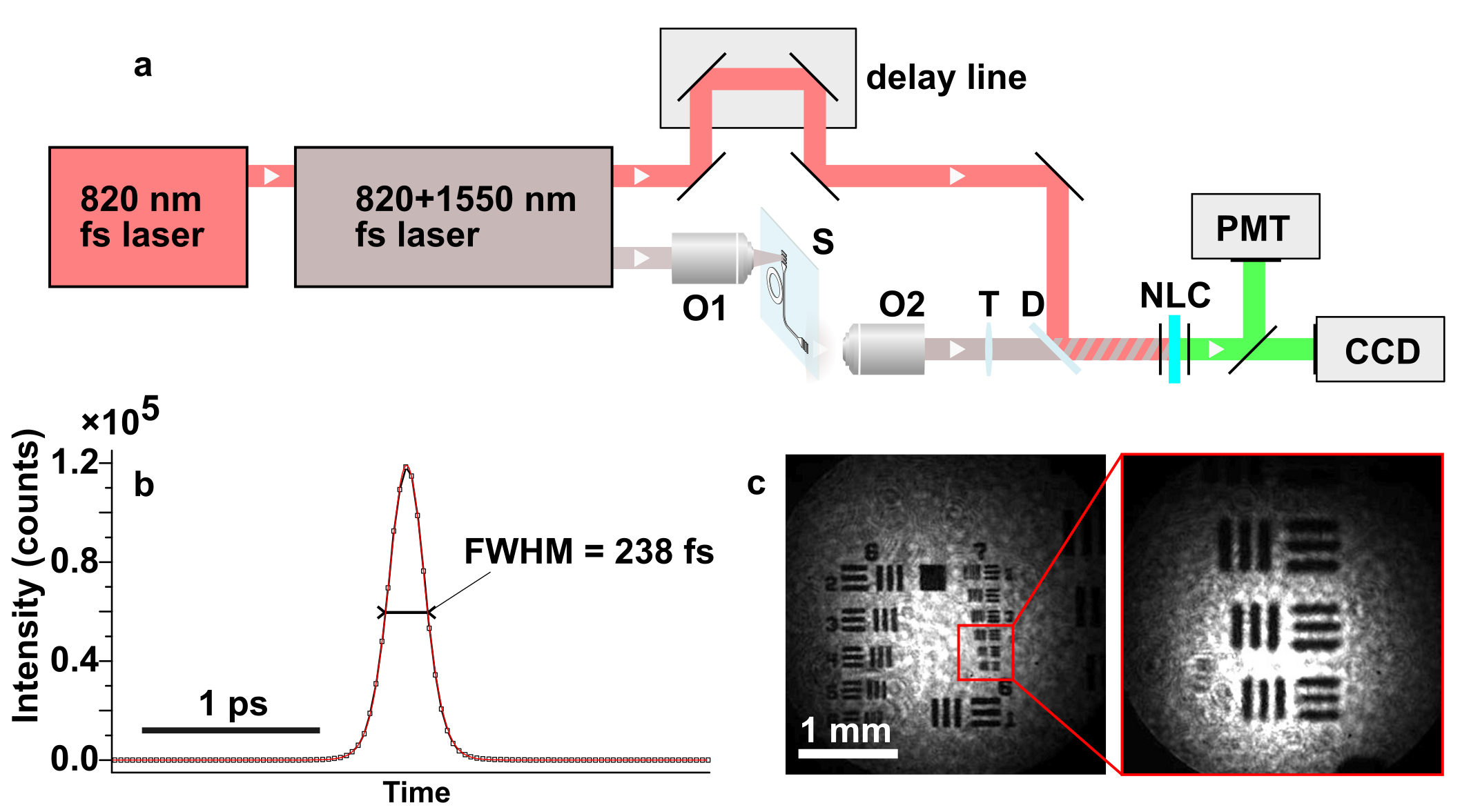}
\caption{(a) Experimental setup of the apparatus (O1 - excitation objective, S - sample, O2 - collection objective, T - tube lens, D - dichroic mirror, NLC - nonlinear crystal, PMT - photomultiplier tube, CCD - charge coupled device). (b) Time resolution retrieved by cross-convolution of the probe and pump beams. (c) Upconverted image of the spatial resolution target USAF 1951, shown at different magnifications, reaching a spatial resolution of \SI{228}{lp\per\milli\meter} with a \num{100}$\times$ objective.}
\label{fig:1}
\end{figure}

The apparatus developed for ultrafast imaging of photonic circuits is reported in Figure \ref{fig:1}. It is based on the principle of optical gating \cite{Trebino1997}: two synchronous femtosecond pulses are made to overlap temporally and spatially in the volume of a nonlinear medium to generate a sum-frequency signal when the phase matching condition is satisfied. The pulses are labeled probe and gate, since the former interacts with the sample while the latter arrives to the nonlinear medium with a preset time delay, defined by a motorized delay line. Any probe signal light reaching the non-linear medium with the same delay of the gate beam results in an upconverted signal at the sum frequency with an intensity proportional to their cross-convolutions, which can then be integrated with a slow detector. 

The setup is based on a previously developed optical gating system \cite{Pattelli2016} but improves significantly on its spatial resolution, exceeding a value of \SI{228}{lp\per\milli\meter} required to resolve features with a characteristic size of about \SI{1}{\micro\meter} (see Figure \ref{fig:1}c). On the other hand, temporal resolution depends on the duration of the pulses that are cross-convoluted (in the order of $\sim$\SI{100}{\femto\second}), but still allows to detect relative time differences in the \si{\femto\second} scale thanks to the combined spatio-temporal detection scheme.

\subsection{Single mode waveguide characterization}
\begin{figure}[ht!]
\centering\includegraphics[width=0.5\textwidth]{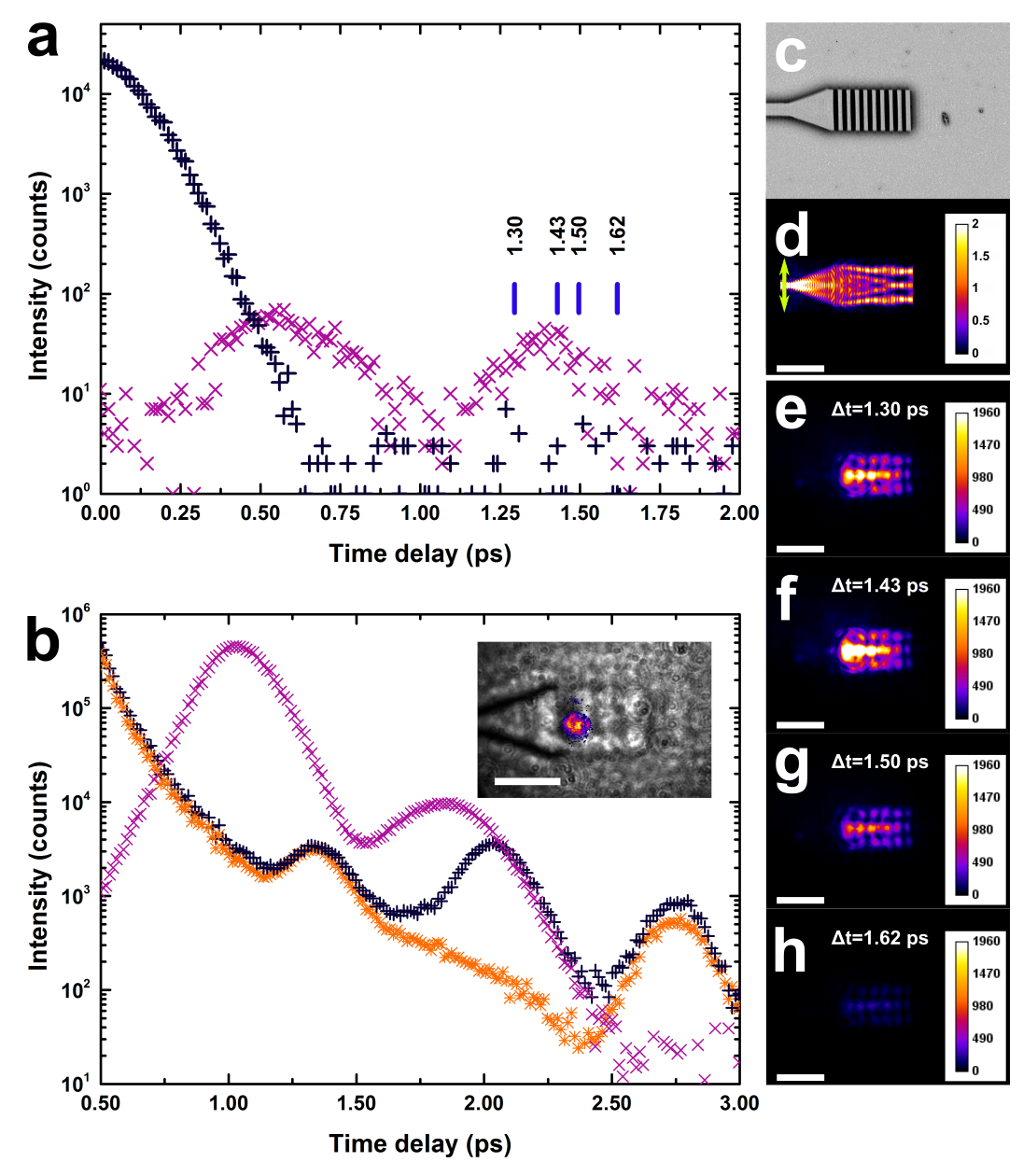}
\caption{Ultra-fast imaging of a single-mode waveguide without defects. The time-resolved characterization reported in panels a and b is referred to two waveguides with different length. (a) Time traces collected from input (black) and output (magenta) couplers. Blue notches correspond to the time instants at which panels e-h have been recorded. (b) Comparison between signals collected from the input and output couplers: at the input coupler (black curve) reflection from the waveguide end is detected at a time delay that is exactly twice the time the pulse needed to propagate forward in the waveguide (image shown in the inset). The magenta curve indicates the time trace from the output coupler with a peak indicating the pulse arrived at one waveguide end. Orange line shows the reference signal taken from substrate without any photonic structure. (c) SEM image of the output grating. (d) Results of finite element method (FEM) based calculation. (e)-(h) Time-resolved images of light propagating inside the output grating coupler. All scale bars are \SI{10}{\micro\meter}}.
\label{fig:2}
\end{figure}

At first, the setup has been tested to characterize single mode waveguides. The photonic structures presented in this work have been designed for operation at telecom wavelengths and fabricated in a polymeric matrix by 3D lithographic technique of Direct Laser Writing (DLW). Free-space beam propagation is converted into an on-chip guided mode by grating couplers optimized to work at perpendicular incidence for a selected polarization.
Assuming a negligible dispersion in the wavelength range of interest, the effective refractive index of our single-mode waveguide can be estimated by measuring the time delay acquired by the pulse propagating along a waveguide of known length.
Characterization of waveguides of various configuration is summarized in Figure \ref{fig:2}. Figure \ref{fig:2}a shows the time-resolved transmission profile through a single-mode waveguide as collected from the input and output couplers. The signal passing through the input coupler directly into the camera sets a zero-time reference for the light pulse that propagates in the photonic circuit. Based on the waveguide length, the measured pulse time delay at the output coupler of $\Delta t = \SI{1.376}{\pico\second}$ allows us to retrieve an effective refractive index of $n_\text{eff}=\num{1.49}$, in good agreement with the result of numerical calculations. 

Looking more closely at the late time range of a different waveguide configuration (see Figure \ref{fig:2}b), a second peak at the input coupler is clearly visible at a delay that is the double of the time spent propagating along the waveguide, showing that the setup can be operated also in an ODTR configuration.
Furthermore, it is interesting to observe the light dynamics at the output grating coupler (grating pitch $\mathit{\Lambda} = \SI{1050}{\nano\meter}$) tracking its time evolution inside the structure (Figure \ref{fig:2}e-h). The acquired spatio-temporal images show that the extinction length inside the output coupler is shorter than the grating size, confirming the effectiveness of its design. 

\begin{figure}[h!]
\centering\includegraphics[width=\textwidth]{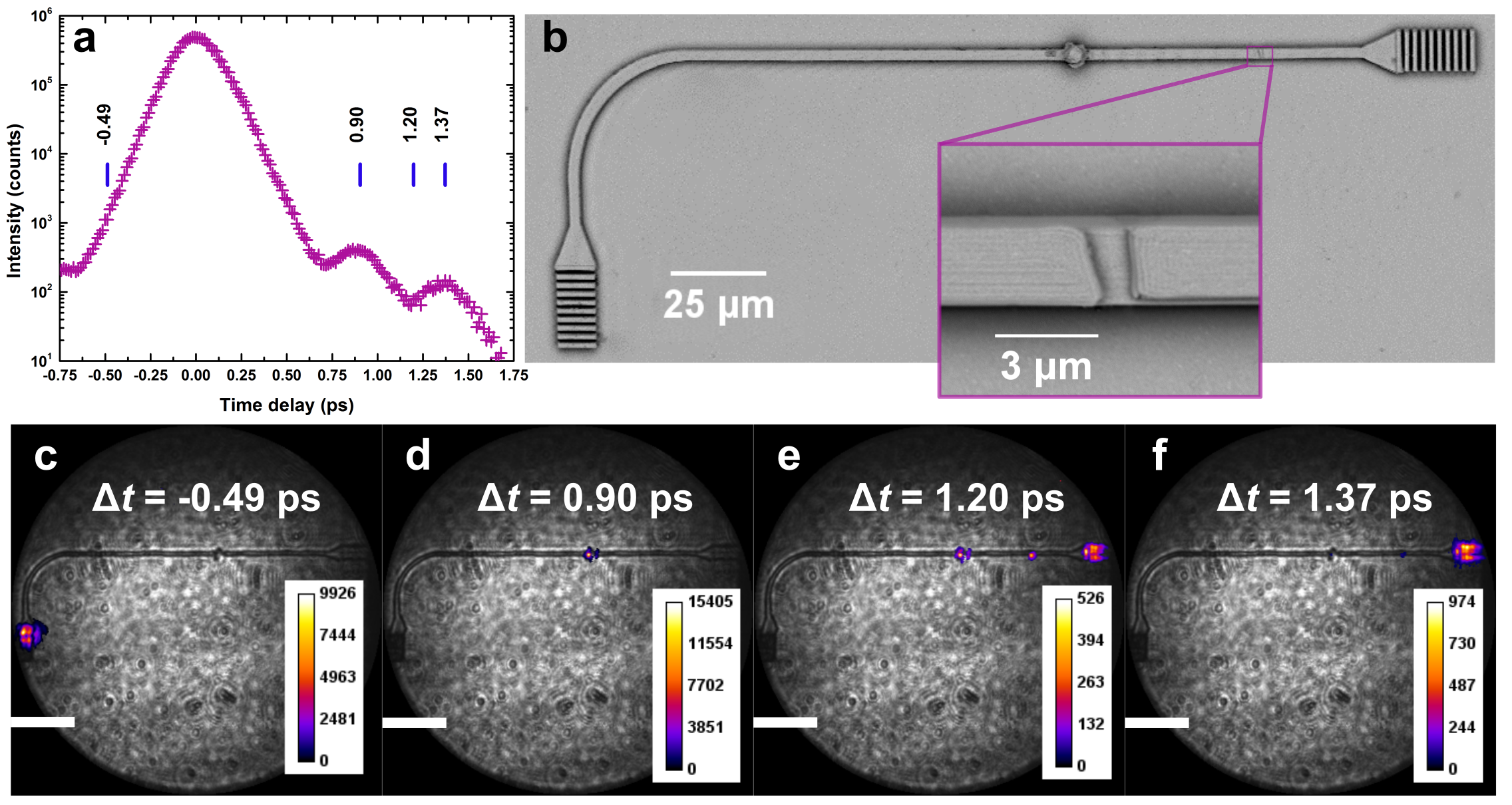}
\caption{Analysis of defects in single mode waveguides with ultra-fast imaging technique. (a) Time-resolved transmission through the waveguide with two defects. The peak at $t=0$ corresponds to the pulse impinging into the structure, the second peak is mainly due to scattering from the first defect and the last peak is the power outcoupled from the waveguide. It is important to note that although scattering from the second defect is not distinguishable in time domain, it can be clearly resolved in spatio-temporal domain (panel e). Blue notches correspond to the images c-f. (b) SEM image of the waveguide under investigation; the inset shows the designed indentation. (c)-(f) Images (see Methods) taken at successive time delays showing the evolution of the light pulse in the waveguide. All scale bars are \SI{50}{\micro\meter}.}
\label{fig:3}
\end{figure}

This first example shows basic diagnostic steps to evaluate the effective refractive index of the mode and track light dynamics at the grating coupler in a simple circuit with no defects. However, different types of imperfections may be introduced accidentally during fabrication. For example, using DLW lithographic patterning, photonic components can be affected by scattering defects due to incidental polymer burning or deviations from the designed geometry because of positioning errors. Similar problems may occur with other fabrication techniques as e-beam lithography. To investigate such issues, a single mode waveguide with two defects was analyzed (Figure \ref{fig:3}).  One defect is a small indentation on the top of the waveguide (Figure \ref{fig:3}b, inset), which was intentionally introduced to slightly perturb the mode. The second defect is an accidental polymer burning occurred during fabrication. As we show in Figure \ref{fig:3}, thanks to the combination of spatial and temporal imaging, the technique allows to discriminate and investigate separately the two scattering events even though they are separated by just \SI{\sim 50}{\micro\meter}, corresponding to a time separation of  \SI{\sim 250}{\femto\second}.
While the short separation between the two events makes them unresolvable in the time domain, the imaging acquisition allows to selectively monitor either defect allowing to address individual time traces. On the other hand, temporal resolution allows to effectively filter out the impinging light pulse which would otherwise dominate over the weak signal scattered by the defects by several orders of magnitude. This example illustrates the complementarity of spatio-temporal information, which cannot be reduced to the sum of spatial and temporal measurements performed separately.

\subsection{Whispering gallery mode resonator characterization}
\begin{figure}[ht]
\centering\includegraphics[width=.5\textwidth]{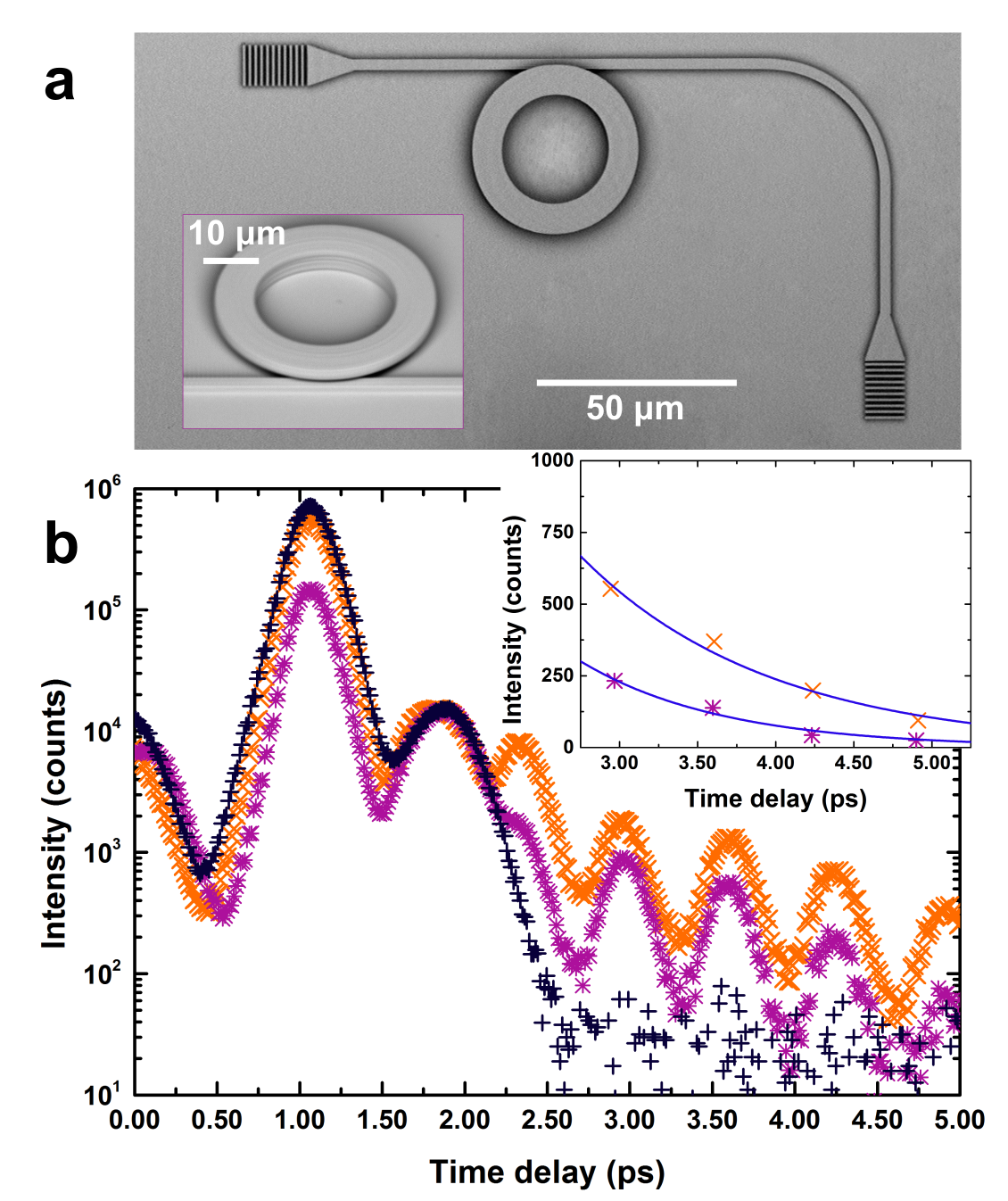}
\caption{Study of vertically coupled whispering gallery mode resonator. (a) SEM image of typical waveguide-resonator photonic circuit. Inset shows the vertical coupling geometry of the structure. (b) Time traces recorded at the output coupler. Black curve corresponds to the case when there is no WGM resonator. The first peak at $t \simeq \SI{1}{\pico\second}$ corresponds to the fraction of light propagating in the bus waveguide (not coupled into the resonator). The peak at $t \simeq \SI{1.75}{\pico\second}$ is a reflection artifact from the substrate (verified by imaging at this time delay). The pulses recorded after $t \simeq \SI{2}{\pico\second}$ correspond to the light coupled into the resonator that made several round trips. Magenta and orange curves correspond to different resonators with different quality factors ($Q_1 = \num{1000}$, $Q_2 = \num{1600}$). Inset shows the exponential decays of the intensity used to extrapolate the photon lifetime.}
\label{fig:4}
\end{figure}

In order to show the applicability of the method to more complex circuits, we study a whispering gallery mode resonator vertically coupled (Figure \ref{fig:4}a) to a single-mode waveguide designed for telecommunication C-band.
Within this structure, light propagation dynamics is governed by multiple factors. The probe pulse is coupled into the waveguide through a grating coupler and propagates through the waveguide reaching the resonator. Depending on waveguide-to-cavity coupling coefficient and the resonant wavelength, a certain fraction of the guided light is coupled into the resonator, while the rest keeps propagating through the waveguide \cite{Ghulinyan2013}. The electromagnetic field that is coupled into the resonator starts to circulate. At each round trip, the resonant cavity mode loses a fraction of power back into the bus waveguide, which depends on the same coupling coefficient (insertion losses), and into free space, which instead depends on the intrinsic quality factor of the cavity \cite{Kippenberg2003}. Light returning to the waveguide carries information about the total quality factor (the sum of the intrinsic and the external ones) of the whispering gallery mode resonator. This results in exponentially decaying, time-equidistant peaks in the time-resolved transmission profile. Compared to traditional cavity ringdown measurement, the high temporal resolution of this setup allows to identify each cavity round trip instead of just the exponential intensity decay \cite{Armani2003}. The exponential decay of the peak intensity is connected to the photon lifetime, characterizing the energy dissipation rate (Figure \ref{fig:4}b, inset). 

\subsection{Multiple port waveguide network characterization}
\begin{figure}[p]
\centering\includegraphics[width=.5\textwidth]{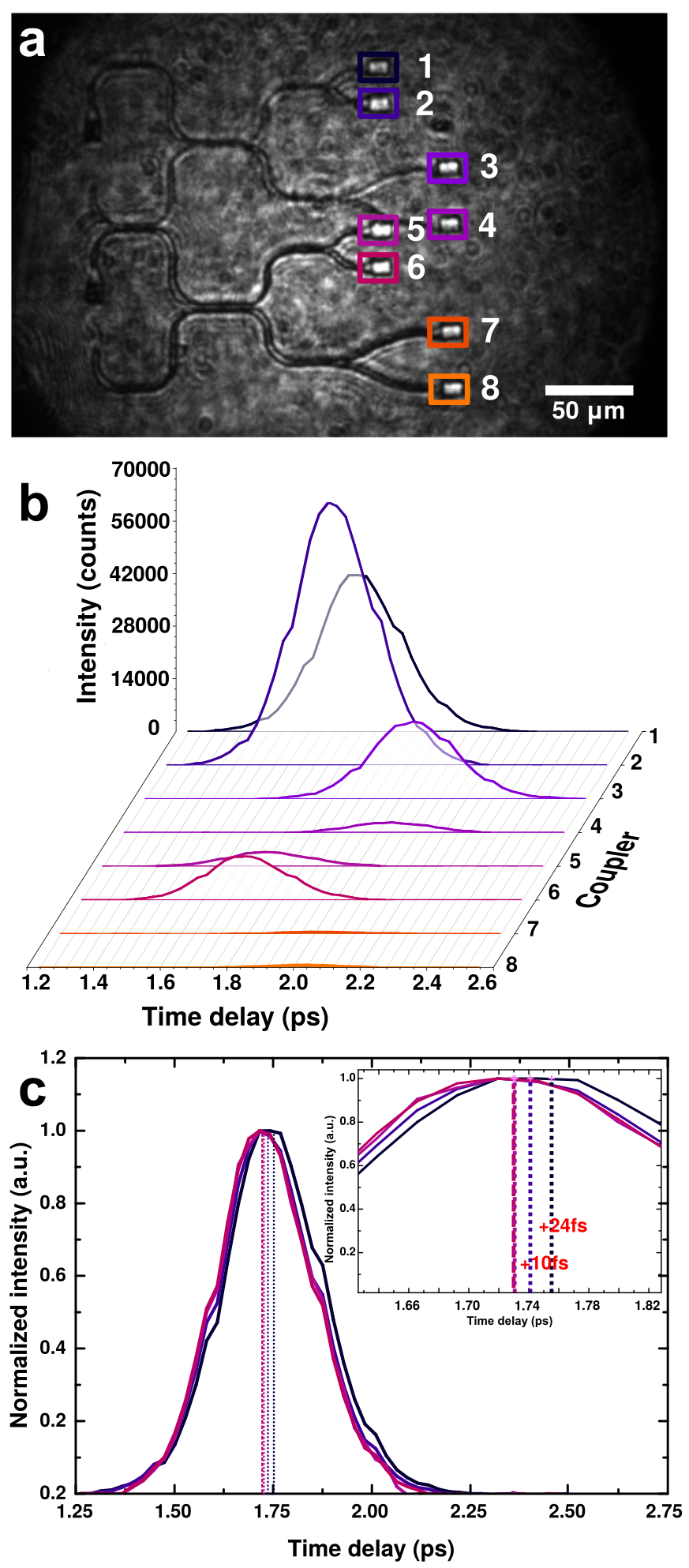}
\caption{Interconnected waveguide network analysis. (a) A wide-field image of the multi-channel inter-coupled waveguide network. Colors and numbers identify different output ports. (b) Time-intensity distribution of the signal from different output ports. Peaks forms two groups depending on the path length. (c) Temporal signal distribution of the four pulses with a time delay of about \SI{1.75}{\pico\second} (1\textsuperscript{st}, 2\textsuperscript{nd}, 5\textsuperscript{th} and 6\textsuperscript{th} output ports). Insets shows the zoom of the four peak centers, illustrating the temporal mismatch of nominally equal paths differing due to fabrication imperfections.}
\label{fig:5}
\end{figure}

Photonic circuits employed in commercial devices usually include multiple waveguides to interconnect and manipulate optical signals. Coupling among parallel light pathways is achieved by directional couplers whose intensity ratio splitting is determined by the coupling length \cite{Kiyat2005}. These highly interconnected circuits represent a promising platform for an array of applications including all-optical computation and artificial intelligence \cite{Shen2017}. A typical system consists of several input and output ports wired through waveguides connected via directional couplers. As directional couplers are extremely sensitive to fabrication imperfections, a structure containing nominally identical parts might still result in an undesired transmission profile even without any visible defect. Analysis of such devices using traditional reflectometry techniques is impractical due to the high number of input-output port combinations that must be addressed individually and the impossibility to uniquely identify branches of the same waveguide.
Here we apply our time-resolved imaging technique to an exemplary waveguide network, consisting of two input couplers, three directional couplers and four Y-connectors resulting in eight output ports (Figure \ref{fig:5}).
The output ports can be organized into two groups according to their nominal path length to the output coupler. Additionally, each pulse experiences at least one event of partial power transfer through a directional coupler and one power split though a Y-connector. It is interesting to follow the light dynamics through the paths that are nominally identical by design (e.g., output port 1 and 2). Thanks to the far-field imaging nature of our technique, the circuit analysis does not require any scanning process and the whole structure can be monitored at once in a wide field acquisition. As we collect a set of time-resolved images of the entire structure, individual transmission curves are retrieved for each output port by spatially integrating the signal in their corresponding regions.

Figure \ref{fig:5}b depicts time-resolved transmission profiles of all output couplers showing the sensitivity of the method to the intensity distribution at the directional couplers and Y beam splitters. Looking at the normalized time traces (Figure \ref{fig:5}c) of light pulses that propagated through nominally equal optical paths (output ports n.\ 1, 2, 5, and 6), we observe an asynchrony in the arrival times. In the inset, we show that light that is out-coupled from grating 5 and 6, arrives exactly at the same time (within \SI{3}{\femto\second} accuracy), while light that comes out from couplers 1 and 2 differs both from the lower branch and among themselves, highlighting the presence of unintentional imperfections that would be otherwise difficult to identify. 
Furthermore, a quantitative characterization of the performances of the circuit components can be performed by integrating in time the outcoupled intensities. In the case of the waveguide network, it is interesting to calculate the splitting ratio of the three directional couplers that are responsible for the signal propagation through the structure. For the investigated device, the performance of the splitters does differ significantly (90:10,  82:18 and  97:3 for the leftmost, the upper and the lower splitter, respectively) despite their nominally identical design.

\section{Methods}
\label{sec:MatMeth}

\subsection{Experimental apparatus}
Ultrafast imaging was performed using femtosecond pulses produced by a Ti:Sapphire mode-locked infrared laser (Tsunami, Spectra Physics, central wavelength $\lambda_\text{gate} = \SI{820}{\nano\meter}$, $\text{FWHM} = \SI{9}{\nano\meter}$), and an optical parametric oscillator (OPO Opal, Spectra Physics, central wavelength $\lambda_\text{probe} = \SI{1550}{\nano\meter}$, $\text{FWHM} = \SI{15}{\nano\meter}$). At the phase-matching condition, sum-frequency generation occurs in a \SI{0.5}{\milli\meter} thick bismuth borate (BiBO) crystal. The gate pulse path is tuned by a linear motorized stage (ThorLabs Inc., ODL300, \SI{300}{\milli\meter} Travel), which can be moved with precision up to \SI{0.5}{\micro\meter} (corresponding to a round-trip delay of $\sim$\SI{3}{\femto\second}). Sum-frequency signal with $\lambda_\text{signal} = \SI{536}{\nano\meter}$ is recorded by a photomultiplier tube (ET Enterprises) or low noise CCD camera (Andor iKon M912). 
Light emerging from the sample is collected by a set of collection optics imaging the spatial distribution of light on the nonlinear crystal facet. The same facet of the crystal is illuminated with an expanded gate beam providing a uniform illumination over its surface to ensure flat upconversion efficiency all over the image.
Polarization of the gate pulse is tuned by a half-wave plate in order to achieve maximal in-coupling. This allows in turn to select the polarization of collected light through the upconversion process by rotating the non-linear crystal. 

\subsection{Image acquisition}
Time-resolved images reported in the paper are overlaid to wide-field illumination images for the sake of clarity. The probe beam is focused (Mitutoyo NIR 10$\times$ NA \num{0.26}) on a target input coupler and the output signal is collected by an infinity-corrected collection objective (Mitutoyo NIR 20$\times$ NA \num{0.4}, 50$\times$ NA \num{0.42} or 100$\times$ NA \num{0.5}). Images at different delay values (shown in color) are recorded for different positions of the translation stage, and superimposed to a zero-delay (grayscale) image recorded with a collimated probe beam illumination. This allows to see the structure through the same set of imaging optics (including the nonlinear crystal) and identify the structural features corresponding to each signal.

\subsection{Sample fabrication}
Investigated devices have been fabricated using a commercial 3D Direct Laser Writing platform (NanoScribe GmbH) based on two-photon polymerization employing a commercial photoresist (Ip-Dip from Nanoscribe GmbH, $n=\num{1.53}$ \cite{Gissibl2017}). The final polymeric structures are not limited to planar geometries but can be patterned in the whole 3D space in a point-by-point polymerization process whose resolution is determined by the voxel dimension, which has an ellipsoidal shape with a minor axis length of \SI{120}{\nano\meter} and a major axis of \SI{250}{\nano\meter}). 
Using a glass substrate with a refractive index (fused silica $n=\num{1.444}$ at \SI{1550}{\nano\meter}) that is lower than that of the photoresist, a ridge waveguide configuration provides good mode confinement in single mode waveguides. Light coupling into the waveguide is enabled by grating couplers that convert a free-space propagating probe pulse to an on-chip guided mode \cite{Nocentini2018}.

\subsection{Numerical modeling}
Finite element based calculations were performed using a commercial software (Comsol 4.3), to optimize and evaluate the mode propagation inside the photonic structures. 
Mode analysis calculation was performed to evaluate the effective refractive index of the guided mode within the single mode waveguide while grating coupler design has been optimized in order to maximize free-space to on-chip light propagation (frequency domain calculation). 

\section{Conclusions}
We described a novel technique for functional photonic circuit diagnostics and quality control. This technique merges complementary approaches enabling a detailed characterization of photonic components. Notably, spatio-temporal characterization of photonic circuits gives access to device linear and non-linear characteristics. Although the technique is demonstrated for the common C-band wavelength region, its basic working principle is that of optical gating which can be readily extended to other frequency regions. Measurements are performed in a non-contact, far-field fashion making this method compatible with 3D or vertically stacked structures that would not be suitable for a scanning near-field investigation. Besides, the technique can be generalized to work in reflection mode, lifting the requirement for transparent substrate. Finally, the technique does not rely on any specific material characteristics, and is therefore applicable also to silicon, hybrid or polymer photonic platforms.In this respect, our proposed technique fills the lack of available diagnostics approaches for complex multi-branched integrated circuits, while at the same time offering a defect positioning resolution that is significantly superior to the state-of-the-art demonstrated so far over linear waveguides using OFDR.

Remarkably, the diagnostics potential of our technique extends not just to geometrical defects close to the optical diffraction limit, but it can detect also slight variations from the intended design of investigated photonic circuits. Indeed, there can be imperfections that are undetectable even under scanning electron microscopy inspection (e.g., regarding the material density or the presence of buried defects) and yet affect the functionality of the photonic device. On the other hand, devices exhibiting slight geometrical deviations might still perform as intended, in ways that are impossible to predict or model beforehand and that can only be assessed using in situ, functional diagnostic techniques.

\section*{Funding}
This research was supported by Erasmus Mundus Doctorate Program Europhotonics (Grant No. 159224-1-2009-1-FR-ERA MUNDUS-EMJD) and ERC Advanced Grant n.\ 291349.

\section*{Acknowledgements}
We thank Simone Zanotto for fruitful discussions and Jeroen van den Brink for technical assistance during construction of the apparatus.

\section*{Disclosures}
The Authors declare financial competing interests related to a pending patent: Italian priority n.\ 102018000008647 co-owned by Universit\`a degli Studi di Firenze and Laboratorio Europeo di Spettroscopia Non-Lineare (P). 
\\[1cm]
See Supplement 1 for supporting content.

\bibliography{main}

\end{document}


\maketitle
\begin{abstract}

\textit{This document is the unedited Author\textquotesingle s version of a Submitted Work that was subsequently accepted for publication in ACS Photonics, copyright \textcopyright American Chemical Society after peer review. The final edited and published work is available at  \href{https://dx.doi.org/10.1021/acsphotonics.0c00271}{ DOI: 10.1021/acsphotonics.0c00271}.}
\end{abstract}{}
\section{Calculation of quality factors}

Quality factor of the resonator was estimated by measuring the time-resolved transmission through the waveguide coupled to the resonator. This allows to resolve in time each round-trip of a pulse inside the resonator and thus calculate $Q$-factors similarly to what is done in cavity-ringdown spectroscopy.  Each single peak that corresponds to the effect of the WGM resonator is approximated with a Gaussian fit to calculate the area below it. Then using this information together with the peak delays, a single exponential decay fit is performed.
\begin{equation}
I = C+I_0 \mathrm{e}^{-t/\tau}
\end{equation}

where $I$ is the intensity registered by detector, $C$ is a fitting constant, $I_0$ is the peak intensity, $t$ is time and $\tau$ is the photon cavity lifetime.

Since the energy stored in the resonator is decaying to approach zero, the $C$ parameter is set to zero. The decay time and quality factor $Q$ is therefore estimated as
\begin{equation}
Q = \frac{2\pi c\tau}{\lambda}
\end{equation}

where $c$ is the speed of light and $\lambda$ is the wavelength for which the calculation is performed.

\section{Numerical simulation}
\begin{figure}[htb]
\centering
\includegraphics[width=0.5\textwidth]{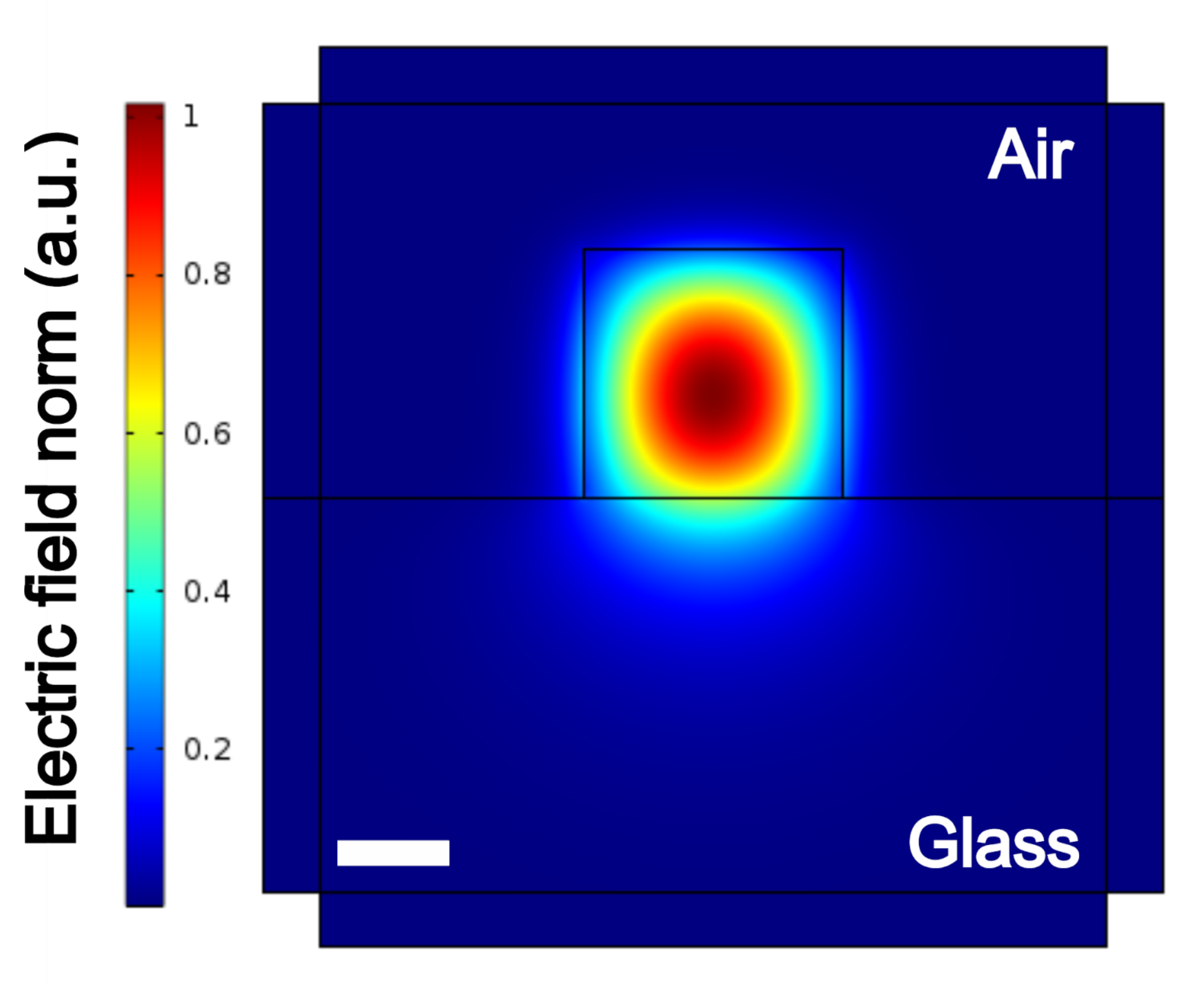}
\caption{Electric field distribution in a single mode waveguide modelled with mode solver. Effective refractive index of the mode is \num{1.486}. The scale bar is \SI{1}{\micro\meter}.}
\label{fig:1}
\end{figure}

The calculation of the effective refractive indices of guided modes was performed with Finite Element Method (FEM) software Comsol 4.3 (mode solver). The electric field distribution for one of the designed waveguide modes is shown in \ref{fig:1}.

The value of the effective refractive index is used further for modeling electric field distribution in Bragg couplers. Moreover, setting the source as boundary mode port (Figure 2d) in 2D geometry requires knowledge of the effective refractive index of the mode.